\documentclass[twocolumn,preprintnumbers,superscriptaddress,prl,nofootinbib,longbibliography,amsmath,amssymb]{revtex4-1}

\usepackage{graphicx}
\usepackage{dcolumn}
\usepackage{bm}
\usepackage{color}
\usepackage{ulem}
\usepackage{gensymb}
\usepackage{braket}
\usepackage{amsmath}
\usepackage[percent]{overpic}

\begin{document}

\title{Orbital degree of freedom in high entropy oxides}

\author{Jiaqiang Yan}
\email{yanj@ornl.gov}
\affiliation{Materials Science and Technology Division, Oak Ridge National Laboratory, Oak Ridge, Tennessee 37831, USA}

\author{Abinash Kumar}
\affiliation{Materials Science and Technology Division, Oak Ridge National Laboratory, Oak Ridge, Tennessee 37831, USA}

\author{Miaofang Chi}
\affiliation{Materials Science and Technology Division, Oak Ridge National Laboratory, Oak Ridge, Tennessee 37831, USA}

\author{Matthew Brahlek}
\affiliation{Materials Science and Technology Division, Oak Ridge National Laboratory, Oak Ridge, Tennessee 37831, USA}

\author{Thomas Z Ward}
\affiliation{Materials Science and Technology Division, Oak Ridge National Laboratory, Oak Ridge, Tennessee 37831, USA}

\author{Michael McGuire}
\affiliation{Materials Science and Technology Division, Oak Ridge National Laboratory, Oak Ridge, Tennessee 37831, USA}

\date{\today}

\begin{abstract}

The spin, charge, and lattice degrees of freedom and their interplay in high entropy oxides were intensively investigated in recent years. However, how the orbital degree of freedom is affected by the extreme disorder in high entropy oxides has not been studied. In this work, using perovskite structured \textit{R}VO$_3$ as a materials playground, we report how the disorder arising from mixing different rare earth ions at the rare earth site affects the orbital ordering of V$^{3+}$ t$_{2g}$-electrons. Since each member of \textit{R}VO$_3$ (\textit{R}\,=\,rare earth and Y) crystallizes into the same orthorhombic \textit{Pbnm} structure, the configurational entropy should not be critical for the stability of (\textit{R}$_1$,...,\textit{R}$_n$)VO$_3$. The spin and orbital ordering was studied by measuring magnetic properties and specific heat of single crystals. Rather than the number and type of rare earth ions, the average and variance of ionic radius are the key factors determining the spin and orbital order in  (\textit{R}$_1$,...,\textit{R}$_n$)VO$_3$. When the size variance is small, the average ionic radius takes precedence in dictating spin and orbital order. Increasing size variance suppresses the G-type orbital order (G-OO) and C-type antiferromagnetic order (C-AF) but favors the C-OO/G-AF state and spin-orbital entanglement. These findings suggest that the extreme disorder introduced by mixing multiple rare earth ions in high entropy perovskites might be employed to preserve the orbital degree of freedom to near the magnetic order, which is necessary for the electronic driven orbital ordering in a Kugel-Khomskii compound.

\end{abstract}

\maketitle

\section{Introduction}

Inspired by the concept of high entropy alloys\cite{yeh2004nanostructured,cantor2004microstructural}, high entropy oxides incorporate multiple cationic species in equimolar or nearly equimolar ratios in their crystal lattice. The pioneering work on entropy stabilized oxides in 2015 sparked immense research excitement within the scientific community, as it provides a pathway to explore novel emergent phenomena and customize material properties within an expansive compositional space\cite{rost2015entropy}. This breakthrough has stimulated extensive investigations aimed at uncovering the unique characteristics and potential applications of high entropy oxides\cite{oses2020high}. Extensive research efforts have been dedicated to studying the spin, charge, and lattice degrees of freedom  as well as their interplay in high entropy transition metal oxides\cite{salian2022entropy,kotsonis2023high,witte2019high,meisenheimer2020oxides,xiong2022low,sarkar2021magnetic,sharma2023large,kumar2023magnetic,farhan2022weak,li2023preparation,jin2023robust,sarkar2023high,su2022direct,kumar2023thermoelectric,mazza2022designing}. However, the influence of chemical disorder on the orbital degree of freedom in these materials remains unexplored.

Perovskite-structured rare earth transition metal oxides provide an excellent platform to address this question. While mixing different transition metal ions at the transition metal site can introduce complexities related to spin and valence states, utilizing different rare earth ions at the rare earth site proves to be a more viable option. For instance, investigating the effects on the orbital ordering of t$_{2g}$ electrons can be accomplished by studying materials such as (\textit{R}$_1$,...,\textit{R}$_n$)VO$_3$, while insights into the impact on the orbital degree of freedom of the e$_g$ electrons can be gained by studying (\textit{R}$_1$,...,\textit{R}$_n$)MnO$_3$ perovskites. This raises the question of why and how the composition complexity and hence lattice disorder at the rare earth site should affect the orbital order at the transition metal site.

It is well-established in perovskites that the covalency between rare earth ions and oxygen can give rise to the interactions between Jahn-Teller and GdFeO$_3$ distortions, consequently affecting the orbital order\cite{mizokawa1999interplay}. The direction of the rare earth ions's shift is intimately connected to the specific orbital order. In a high entropy oxide  (\textit{R}$_1$,...,\textit{R}$_n$)\textit{M}O$_3$ (\textit{M}\,=\,transition metal), the magnitude of the local lattice distortion is not uniform even if all \textit{R}$_i$\textit{M}O$_3$ members have the same type of orbital order at low temperatures. When the constitutional \textit{R}$_i$MO$_3$ members have different types of orbital order  at low temperatures, both the magnitude and direction of the rare earth ions' shift can become frustrated. In both cases, it is expected that the long range orbital order is disturbed by the extreme lattice disorder at the rare earth site. Therefore, it becomes intriguing to explore the mechanisms through which the orbital order is suppressed and whether the chemical disorder at the rare earth site can lead to the emergence of novel orbital states. Investigating these aspects would provide valuable insights into the interplay between chemical disorder and orbital degree of freedom, opening up new avenues for understanding and manipulating the orbital properties in high entropy oxides.

\begin{figure*} \centering \includegraphics [width = 0.9\textwidth] {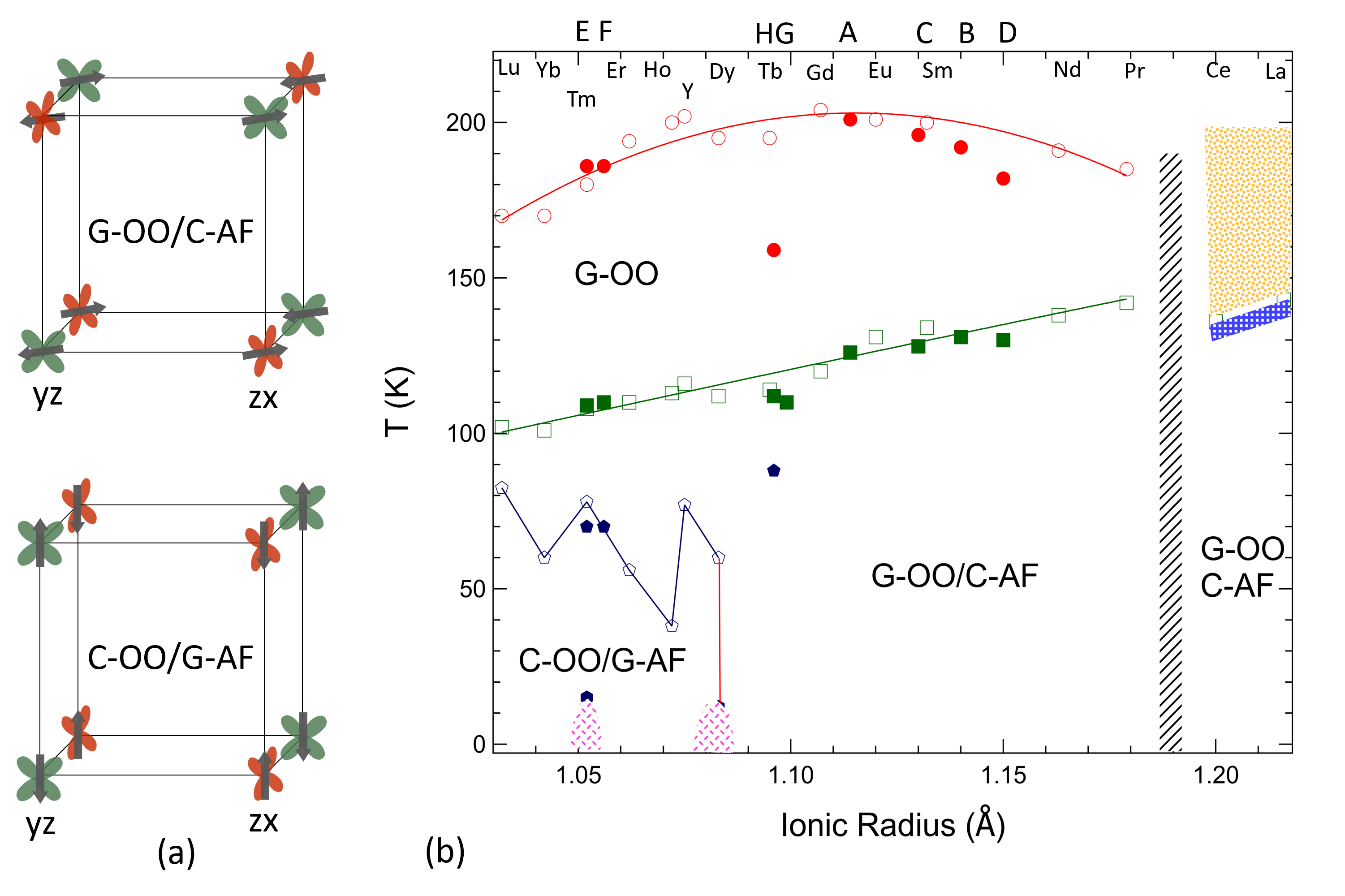}
\caption{(color online) Phase diagram for spin and orbital ordering in \textit{R}VO$_3$ perovskites.  (a) Schematic picture for (upper) G-type orbital order (G-OO) and C-type antiferromagnetic order (C-AF), (lower) C-type orbital order (C-OO) and G-type antiferromagnetic order (G-AF). (b) Spin and orbital ordering in \textit{R}VO$_3$ (open symbols) and (\textit{R}$_1$,...,\textit{R}$_n$)VO$_3$ (closed symbols). Circles, squares, and diamonds denote orbital ordering, magnetic ordering, and orbital flipping, respectively. Data for \textit{R}VO$_3$ were taken from Refs [\citenum{miyasaka2003spin,miyasaka2007magnetic,tung2007magnetization,reehuis2011structural,tung2008spin}].
The yellow dashed region in the temperature range 140\,K-200\,K shows the spin-orbital entangled state for LaVO$_3$ and CeVO$_3$ \cite{zhou2008frustrated, yan2019lattice}. The blue dashed region illustrates the presence of the C-OO/G-AF state in the narrow temperature range between the magnetic order and the first order structure transition\cite{yan2019lattice}. The black dashed stripe near the ionic radius of 1.19$\AA$ highlights the difference between PrVO$_3$ and CeVO$_3$. The discrepancy in literature on the spin and orbital ordering of CeVO$_3$ may result from the nonstoichiometry \cite{reehuis2008crystal,tung2008spin,yan2004unusually,munoz2003crystallographic}. The dashed region around Tm and Dy illustrates the low temperature G-OO/C-AF state resulting from the interaction between \textit{R} and V magnetic moments\cite{ritter2016crystallographic,sarkar2015successive,miyasaka2007magnetic,fujioka2010critical,reehuis2011structural}. The letters A to H are positioned according to the average ionic radius of each composition studied in the current work. The ionic radius of rare earth ions with a valence state of 3+ and coordination number of 9 was used since it is the highest coordination number available for all tabulated rare earth ions\cite{shannon1969effective}.}
\label{PhaseDiagram-1}
\end{figure*}

\textit{R}VO$_3$ (\textit{R}=rare earth and Y) perovskites, in which V$^{3+}$ has an electronic configuration of t$_{2g}^2$, show complex spin and orbital orders below room temperature and have been an ideal materials playground for the study of intricate coupling between spin, orbital, and lattice degrees of freedom  of t$_{2g}$ electrons.\cite{miyasaka2003spin} For those two t$_{2g}$ electrons, one takes \textit{xy} orbital, and the other one stays at the degenerate \textit{yz/zx} orbitals. At low temperatures, two types of orbital ordering as shown in Fig.\,\ref{PhaseDiagram-1}(a) are found in \textit{R}VO$_3$ perovskites. In the ab plane of both types of orbital order, \textit{yz} and \textit{zx} orbitals are occupied in a checkerboard pattern. Along \textit{c}-axis, when the stacking is in phase, it gives the so-called C-type orbital order (C-OO); while the out-of-phase stacking results in the so-called G-type orbital order (G-OO). Following Goodenough-Kanamori rules\citep{goodenough1955theory, kanamori1959superexchange}, the corresponding magnetic order is G-type (G-AF) for C-OO and C-type (C-AF) for G-OO, respectively. Figure\,\ref{PhaseDiagram-1}(b) shows the phase diagram for the spin and orbital order in vanadate perovskites. Not all members exhibit the same sequence of spin and orbital order when cooling from room temperature. For \textit{R}=Pr,..., Lu, G-OO occurs around T$_{OO}$=200\,K, then magnetic order around T$_N$=110\,K. For \textit{R}=Dy, ..., Lu and Y, a first order transition occurs around T$_{CG}$=70\,K leading to C-OO/G-AF ground state. Both LaVO$_3$ and CeVO$_3$ have G-OO/C-AF ground state, the same as other members from Tb to Pr. However, instead of an orbital order around 200\,K, a spin-orbital entangled state was suggested in the temperature range T$_N$-200\,K\cite{zhou2008frustrated, yan2019lattice}. A recent theoretical study also suggested LaVO$_3$ as a true Kugel-Khomskii compound\cite{zhang2022lavo}. It should be noted that the competition between the G-OO state and the spin-orbital entangled state is a general phenomenon in \textit{R}VO$_3$ perovskites and this competition can be sensitive to the lattice distortion and nonstoichiometry\cite{yan2019lattice}.

In order to study how the extreme disorder in high entropy oxides affects the orbital degree of freedom, we studied the spin and orbital order in eight different compositions of (\textit{R}$_1$,...,\textit{R}$_n$)VO$_3$. These 8 compositions can be categorized into three groups. Compositions for Group I have rare earth ions \textit{R}= La,...,Tb, and each individual \textit{R}VO$_3$ member has the same G-OO/C-AF ground state. Group II includes compositions with  \textit{R}= Dy,..., Lu and Y, and each individual \textit{R}VO$_3$ member has the C-OO/G-AF ground state. Group III include compositions with \textit{R}=Y, La, ..., Lu and the constituent \textit{R}VO$_3$ can have different spin and orbital ground states. Mixing different rare earth ions modifies the average ions radius and induces size variance at the rare earth site. The rare earth ions were selected in each composition to purposely tune either the average ionic radius or the size variance. The average ionic radius and size variance for all 8 compositions are listed in Table 1. The spin and orbital orderings were studied by measuring magnetic properties and specific heat of single crystal samples. We found the primary factors influencing spin and orbital ordering in (\textit{R}$_1$,...,\textit{R}$_n$)VO$_3$ are the average and variance of ionic radius, rather than the number and specific type of rare earth ions incorporated in the lattice. The results indicate that the extreme disorder in high entropy oxides might be employed to maintain the orbital degree of freedom to near the magnetic order, which is necessary for the electronic driven orbital ordering originally proposed by Kugel and Khomskii in early 1970s\cite{kugel1973crystal}.

\section{Experimental Details}

Single crystals were grown using an optical floating zone furnace in a procedure described previously\cite{yan2004unusually}.  A slow rotating rate of $<$10\,rpm for the lower shaft was found to help stabilize the growth. A growth rate of 6\,mm/h was used for all compositions. The room-temperature x-ray diffraction experiment on powder from pulverized single crystals was performed using a PANalytical X’Pert Pro MPD diffractometer to confirm the phase purity of the crystals and determine their lattice parameters. The compositions of the crystals were measured using energy dispersive spectroscopy (EDS) with a Hitachi TM3000
scanning electron microscope and Bruker Quantax70 spectrometer. Scanning transmission electron microscopy (STEM) imaging and Energy dispersive X-ray spectroscopy (EDS)  were performed on sample E (see Table I for composition) with a probe-corrected JEOL NEOARM operated at 200 KV with a probe convergence semi-angle of 28 mrad. Elemental maps were generated using L peaks for Ho, Er, Tm, Yb, and Lu and K peak for V. The temperature dependence of magnetization was measured with a Magnetic Property Measurement System (Quantum Design, QD) in the temperature interval  2 K\,$\leq$T $\leq$300\,K. Specific heat was measured with a QD Physical Property Measurement System (PPMS) in the temperature interval 2\,K\,$\leq$\,x\,$\leq$\,200\,K. Magnetic and specific heat measurements determine the spin and orbital ordering transition temperatures. We then deduce the nature of the observed transitions by referencing existing knowledge on \textit{R}VO$_3$ perovskites shown in the phase diagram (Fig. 1b).

\begin{table*}[tb]
\centering
\caption{Summary of the high entropy oxides studied in this work. To calculate the average ionic radius and size variance, the ionic radius of rare earth ions with a valence state of 3+ and coordination number of 9 was used
since it is the highest coordination number available for all tabulated rare earth ions\cite{shannon1969effective}. The ordering temperatures were determined from the specific heat data as described in Fig. S5 of Supplementary Material. Also listed are the effective moments,  $\mu_{eff}$(Curie-Weiss fitting/expected), obtained from Curie-Weiss fitting of the magnetization in the paramagnetic state and from the calculation assuming the nominal composition.}
\begin{tabular}{c|c|c|c|c|c|c|c}
  \hline
  \hline
  Samples    &      Rare earth ions    &      average IR     &      variance       & $\mu_{eff}$ &  T$_N$ (K)&  T$_{OO}$ (K) & T$_{CG}$\\
\hline
 A   &      (SmEuGdTb)$_{0.25}$   & 1.114  &  1.9$\times$10$^{-4}$  & 6.90/7.18     &  126 & 201 &* \\
 B   &      (PrNdSmEuGd)$_{0.2}$   & 1.14  &  7.2$\times$10$^{-4}$  & 5.09/5.72     &  131 & 192 &*\\
 C   &      (PrNdSmEuGdTb)$_{0.167}$   & 1.13  &  8.8$\times$10$^{-4}$  & 6.22/6.41     &  128 & 196 &*\\
 D   &      (LaNdSmEuGd)$_{0.2}$   & 1.15  &  15$\times$10$^{-4}$  & 4.84/5.17     &  130 & 182 &*\\
\hline
 E   &      (HoErTmYbLu)$_{0.2}$   & 1.052  &  2.0$\times$10$^{-4}$  & 8.03/7.85    &  109 & 186 &70\\
 F &      (YHoErTmYbLu)$_{0.167}$   & 1.056  &  2.4$\times$10$^{-4}$  & 7.43/7.63     &  110 & 186 &70\\
\hline
 G   &      (PrSmTmLu)$_{0.25}$   & 1.099  &  35$\times$10$^{-4}$  & 5.07/5.22     &  110 & * &* \\
 H   &      (PrSmGdHoTmLu)$_{0.167}$   & 1.096  &  25$\times$10$^{-4}$  & 7.01/6.94     &  112 & 159 &88\\
 \hline
 * transition doesn't exist.
 \end{tabular}
\label{tab:summary}
\end{table*}

\section{Results}

\subsection{Crystal characterization by X-ray and EDS}

\begin{figure} \centering \includegraphics [width = 0.47\textwidth] {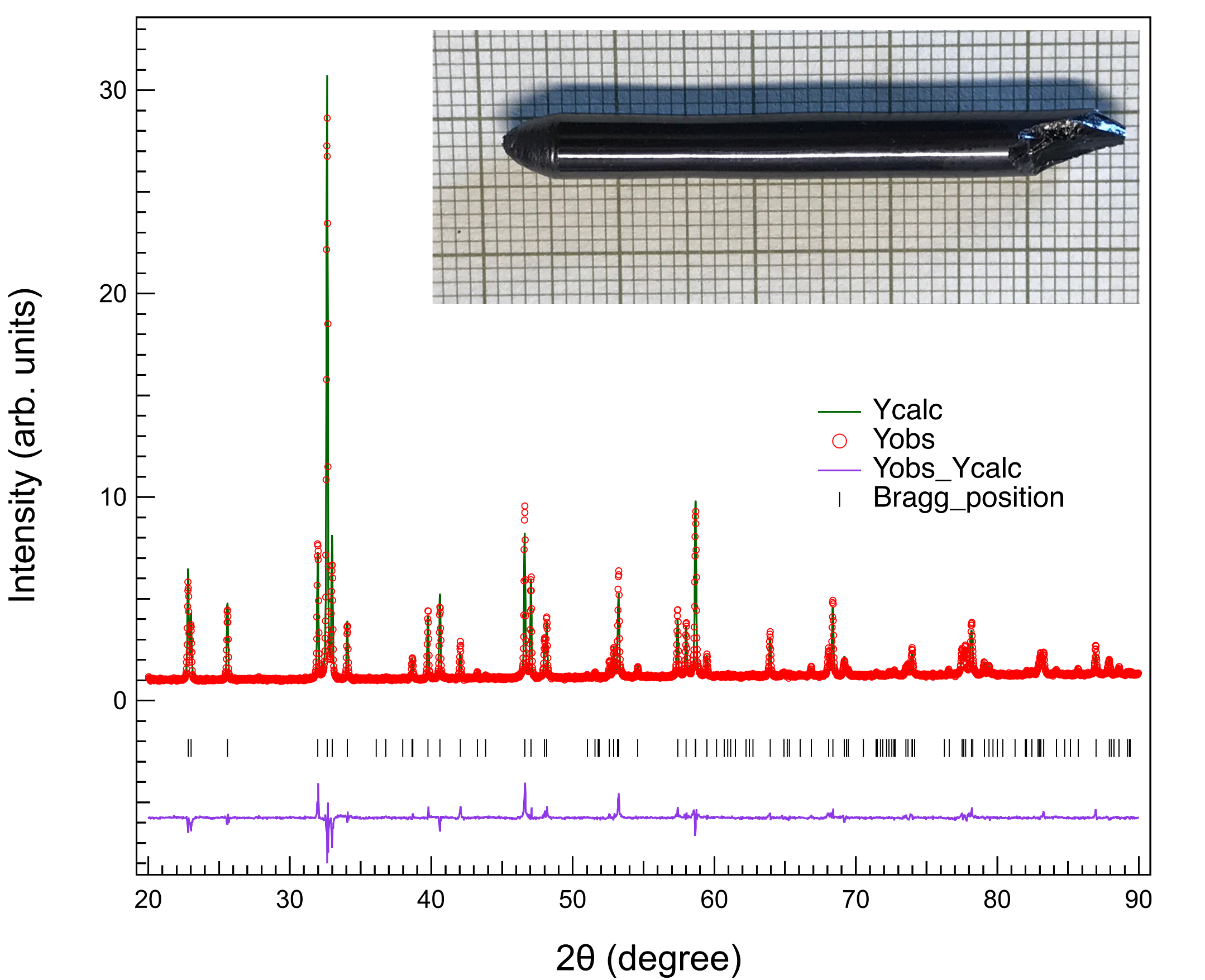}
\caption{(color online) Rietveld refinement of the x-ray powder diffraction pattern of sample D collected at room temperature. Inset shows the single crystal grown by floating zone technique.}
\label{XRD-1}
\end{figure}

As shown in the inset of Fig.\,\ref{XRD-1}, single crystals of around 5\,mm in diameter and several centimeter in length can be obtained by the floating zone technique.  All compositions were confirmed to be single phase by room temperature x-ray powder diffraction. All reflections can be indexed in \textit{Pbnm} symmetry. The lattice parameters of each composition were obtained using Rietveld refinement (Fig. S1, and S2) and shown in Fig. S3 of the Supplemental Material. The ionic radius dependence of the lattice parameters follow that of \textit{R}VO$_3$, indicating lattice parameters are determined by the average ionic radius for   (\textit{R}$_1$,...,\textit{R}$_n$)VO$_3$.  Figure\,\ref{XRD-1} shows the diffraction pattern and the Rietveld refinement of sample D with the composition of (La$_{0.2}$Nd$_{0.2}$Sm$_{0.2}$Eu$_{0.2}$Gd$_{0.2}$)VO$_3$. To confirm the chemical homogeneity of the crystals, we checked with EDS three slices cut from three different positions of the single crystal  of sample D. The equal molar ratio of different rare earth ions was confirmed in all three slices. We also investigated the chemical homogeneity by measuring  EDS using a focused probe in scanning transmission electron microscopy (STEM) mode. The results for sample E was shown in Fig.\,S4 of the Supplemental Material illustrating a uniform chemical distribution among Ho, Er, Tm, Yb and Lu. These sample characterizations suggest that homogeneous single crystals of high entropy perovskites can be grown by floating zone technique. These high quality single crystals enable the investigation of how the lattice disorder affects the orbital degree of freedom as presented below.

\subsection{Spin and orbital order of Group I: (\textit{R}$_1$,...,\textit{R}$_n$)VO$_3$ with \textit{R}\,=\,La, Ce, ..., Tb}

First, we would speculate about possible spin and orbital ordering transitions by referencing the phase diagram shown in Fig.\,\ref{PhaseDiagram-1}. For all four compositions in Group I, the average ionic radius is in the range of 1.114-1.15$\AA$. Assuming the average ionic radius determines the spin and orbital order,  one would expect  G-type orbital order around 200\,K and C-type magnetic order around 120\,K resulting in the G-OO/C-AF ground state for all four compositions. As shown in Fig.S6, a well defined anomaly is observed around 120\,K in the temperature dependence of magnetization, signaling the occurrence of the long range magnetic order. For YVO$_3$ and LuVO$_3$ with nonmagnetic rare earth ions, the orbital order induces a slope change in the reciprocal temperature dependence of magnetic susceptibility\cite{blake2002neutron,yan2005opposing,tung2016magnetic, yan2011spin,yan2007orbital}. In all 8  compositions studied in this work, the magnetic properties in the paramagnetic state are dominated by the magnetic rare earth ions. Thus magnetic measurements will not be able to help identify the orbital order.  Instead, specific heat measurements can be more effective in determining the orbital order transition because lambda type anomalies are expected at both the spin and orbital order transitions\cite{miyasaka2003spin,ren2003orbital,blake2002neutron, tung2016magnetic,tung2007heat,tung2007magnetization,yan2011spin}.

\begin{figure} \centering \includegraphics [width = 0.47\textwidth] {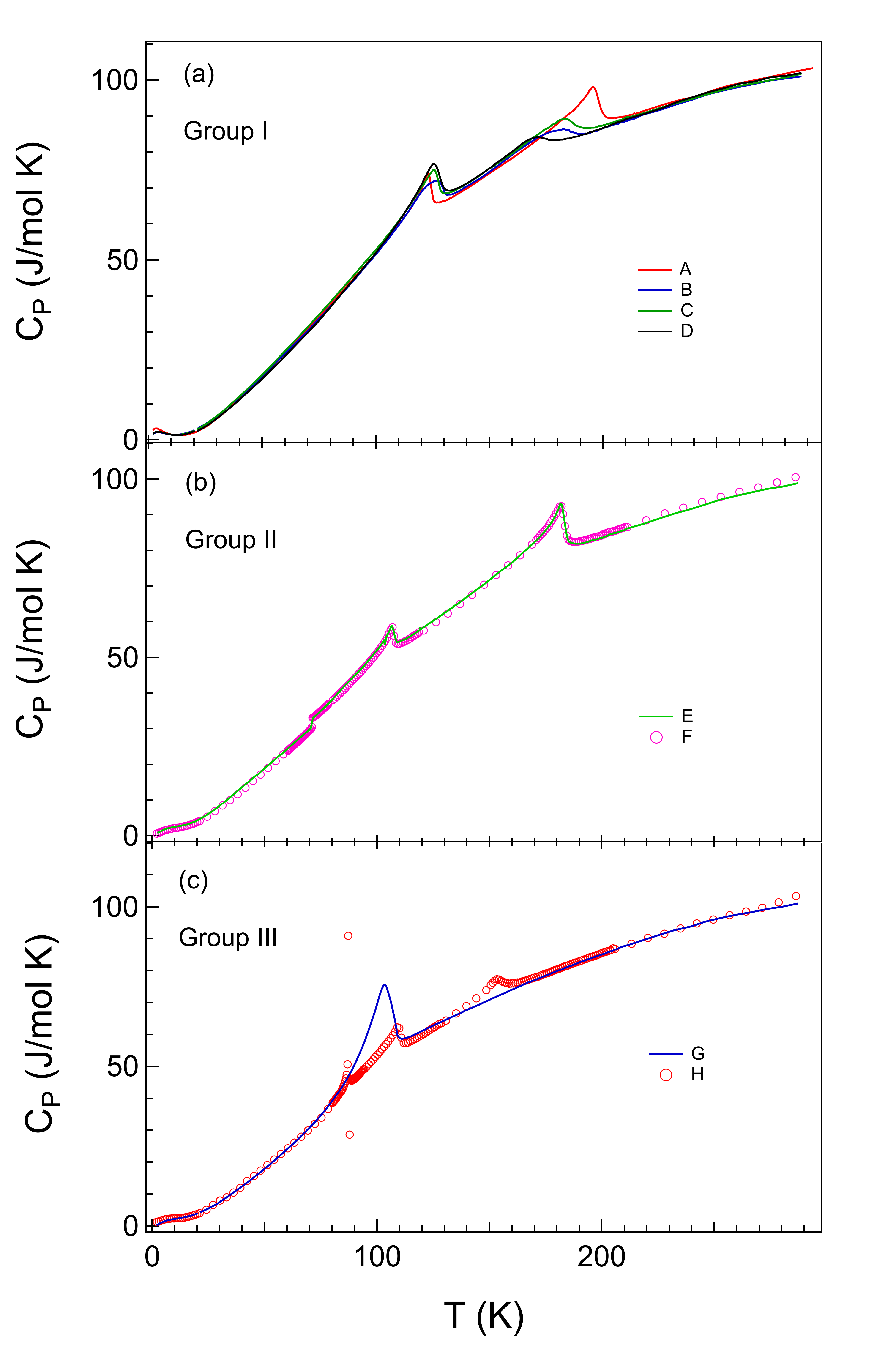}
\caption{(color online) Temperature dependence of specific heat measured in zero magnetic field. (a) Group I compositions. (b) Group II compositions. (c) Group III compositions. }
\label{Cp-1}
\end{figure}

Both G-OO and C-AF transitions are of second order leading to lambda type anomalies in the temperature dependence of specific heat. Figure\,\ref{Cp-1}(a) shows the temperature dependence of  specific heat for all 4 compositions of Group I. As expected, two lambda anomalies can be well resolved in the specific heat curve for each composition. In this work, we define the transition temperature as at which the specific heat starts to arise upon cooling (see Fig.\,S5 in Supplementary Material for details). Based on the sequence of spin and orbital orders in \textit{R}VO$_3$ shown in Fig.\,\ref{PhaseDiagram-1} and previous specific heat studies on \textit{R}VO$_3$ compounds,  it is reasonable to  identify the transition around 200\,K as the orbital ordering and the other one around 120\,K as the magnetic ordering. The temperature dependence of magnetization curves also supports that the low temperature transition is associated with the magnetic order. The transition temperature determined from specific heat data agrees well with that obtained from magnetic measurements.  The spin and orbital ordering temperatures determined from specific heat and magnetic measurements are listed in Table I and also shown in Fig.\,\ref{PhaseDiagram-1}.

From Fig.\,\ref{PhaseDiagram-1}, for \textit{R}VO$_3$ with ionic radius in the range of 1.114-1.15$\AA$, T$_N$ increases while T$_{OO}$ decreases with increasing ionic radius. T$_{N}$ and T$_{OO}$ of group I compositions follow this ionic radius dependence. As discussed later, the slightly suppressed T$_{OO}$  for sample D should be the consequence of its larger size variance.

\subsection{Spin and orbital order of Group II: (\textit{R}$_1$,...,\textit{R}$_n$)VO$_3$ with \textit{R}\,=\,Dy,..., Lu, and Y}

The two compositions in Group II have similar average ionic radius and size variance. The average ionic radius is similar to that of Tm$^{3+}$. From the phase diagram shown in Fig.\,\ref{PhaseDiagram-1}, one would expect to observe three anomalies in specific heat curve for each composition: one lambda anomaly around 180\,K from the G-type orbital order, one lambda anomaly around 110\,K from the C-type magnetic order, and one delta type transition  around T$_{CG}$=70\,K for the first order transition from G-OO/C-AF to C-OO/G-AF upon cooling.

 Specific heat data shown in Fig.\,\ref{Cp-1}(b)  agree with the expectation. The temperature dependence of magnetization (see Fig.S7 in Supplementary Material) also shows well defined anomalies at T$_N$ and T$_{CG}$. The hysteresis observed in magnetization curves measured in heating and cooling modes confirms the first order nature of the transition at 70\,K, consistent with the observation of a delta anomaly in specific heat curve. The transition temperatures determined from specific heat are listed in Table I. As expected, there is little difference between these two compositions. The specific heat jump of 2.6\,J/mol K at T$_{CG}$ is smaller than 3.53\,J/mol K of YVO$_3$\cite{blake2002neutron}, indicating the suppression effect of the size variance. This is supported by an even smaller specific heat jump at T$_{CG}$ of sample H that has a much larger size variance.

\subsection{Spin and orbital order of Group III: (\textit{R}$_1$,...,\textit{R}$_n$)VO$_3$ with \textit{R}\,=\,La, Ce, ..., Lu}

Samples G and H in Group III have a similar average ionic radius that is between those of Tb$^{3+}$ and Gd$^{3+}$. Following the ionic radius dependent spin and orbital ordering shown in Fig.\,\ref{PhaseDiagram-1}, one would expect two lambda anomalies in specific heat for both samples: orbital order around 200\,K and magnetic order around 120\,K. Figure\,\ref{Cp-1}(c) shows the temperature dependence of specific heat. Unexpectedly, sample G has one single lambda anomaly around 110\,K, sample H has two lambda anomalies at 159\,K and 112\,K, and one delta anomaly at 88\,K. This observation suggests that sample H has three transitions as in samples E and F of Group II. Figure\,S8 in Supplementary Material shows the temperature dependence of magnetization, which confirms the magnetic origin of those two transitions at 88\,K and 112\,K for sample H. The step-like change in magnetization at 88\,K suggests this transition is of first order, consistent with the observation of a delta function change in specific heat. Comparison with other \textit{R}VO$_3$ members with similar ionic radius suggests those three transitions are T$_{OO}$=159\,K, T$_N$=112\,K, and T$_{CG}$=88\,K for sample H. As discussed below, the one single transition for sample G comes from the transition between the spin-orbital entangled state and C-OO/G-AF. The large size variance leads to the unexpected spin and orbital ordering transitions for samples G and H.

\section{Discussion}

While the spin and orbital ordering transitions for Group I and II samples suggest that they are determined by the average ionic radius, the unexpected transitions observed for Group III samples highlight the necessity to consider the effect of size variance as well. The ionic radius determines the tolerance factor and the GdFeO$_3$-type lattice distortion that involves the cooperative octahedral site rotations, thus affecting the spin and orbital ordering of the transition metal site. The influence of size variance on the spin and orbital ordering in \textit{R}VO$_3$ perovskites was previously studied in Y$_{1-x}$La$_x$VO$_3$ \cite{yan2007orbital,yan2011spin,yan2019lattice}, Y$_{1-x}$(La$_{0.23}$Lu$_{0.77}$)$_x$VO$_3$\cite{yan2007orbital}, Y$_{1-x}$Eu$_x$VO$_3$ \cite{fujioka2010critical}, Dy$_{1-x}$Tb$_x$VO$_3$\cite{yan2013dy}, and Eu$_{1-x}$(La$_{0.254}$Y$_{0.746}$)$_x$VO$_3$ \cite{fukuta2011effects}. The effects of the size variance can be summarized as below: (1) The size variance tends to suppress the G-type orbital order and C-type magnetic order, but stabilizes the C-OO/G-AF phase; (2) A size variance larger than 3$\times$10$^{-3}$ can induce a spin-orbital entangled state around 200\,K, and upon further cooling this state gives way to the C-OO/G-AF state T$_N$ around 120\,K\cite{yan2019lattice}. This transition seems to be a second order transition from currently available experimental results in literature; (3) For those compositions with an average ions radius near 1.10$\AA$, the size variance can lead to the transition from G-OO/C-AF to C-OO/G-AF upon cooling, which is well demonstrated in the study of Eu$_{1-x}$(La$_{0.254}$Y$_{0.746}$)$_x$VO$_3$ \cite{fukuta2011effects}.

The above results from literature can help understand the unexpected behavior of group III members. Sample G has a variance larger than the threshold of 3$\times$10$^{-3}$, so only one transition is observed at which both orbital and spin order to give the C-OO/G-AF ground state. This also indicates that the spin-orbital entangled state may form in the temperature range from T$_N$ to about 200\,K. This deserves further experimental confirmation. Compared to sample G, sample H has a smaller size variance. This size variance is not large enough to induce the spin-orbital entanglement but large enough to induce a first order transition from G-OO/C-AF to C-OO/G-AF upon cooling through T$_{CG}$. This is similar to what's observed in the study of Eu$_{1-x}$(La$_{0.254}$Y$_{0.746}$)$_x$VO$_3$ \cite{fukuta2011effects}.

The results and analyses described above suggest that the number and type of rare earth ions are not the determining factors for the spin and orbital ordering in high entropy \textit{R}VO$_3$ perovskites.  When the size variance is small, the average ionic radius, or the average structure, determines the spin and orbital ordering temperature. For a fixed average ionic radius, increasing the size variance destabilizes G-OO and C-AF but promotes the C-OO/G-AF state and the spin-orbital entanglement.  This suggests that the disorder from mixing different rare earth ions can tune the competition between spin orbital entangled state and the long range orbital order. This result has important implications on how to design a Kugel-Khomskii compound. When the size variance is above the threshold value of 0.003 in \textit{R}VO$_3$ perovskites, the orbital degree of freedom is maintained even at temperatures near the magnetic order, resulting in the spin-orbital entanglement in the temperature range from  T$_N$ to about 200\,K. One important condition for this to occur is that orbital order should be more sensitive to the size variance than the magnetic order.

This work suggests it would be rather interesting to investigate manganite perovskites (\textit{R}$_1$,...,\textit{R}$_n$)MnO$_3$ with multi-rare earth ions at the same crystallographic site. First, configurational entropy might be employed to  stabilize either the perovskite or the hexagonal phase in compositions with mixing light and heavy rare earth elements. Under ambient pressure, \textit{R}MnO$_3$ members with \textit{R}=La, ..., Tb crystallize in the orthorhombic perovskite structure, those with \textit{R}=Ho,...,Lu and Y stabilize in the hexagonal structure\cite{alonso2000evolution,zhou2006hexagonal}. DyMnO$_3$ can crystallize into either phase depending on the oxygen partial pressure employed in materials synthesis\cite{ivanov2006magnetic}. Perovskite structured \textit{R}MnO$_3$ with \textit{R}=Ho,...,Lu and Y can be synthesized under high pressure and high temperature\cite{alonso2000evolution}. The successful syntheses of single phase orthorhombic (GdLaNdSmY)$_{0.2}$MnO$_3$ and hexagonal (GdTbDyHoErY)$_{0.167}$MnO$_3$ confirm that the configurational entropy can stabilize the perovskite or the hexagonal \textit{R}MnO$_3$  when mixing light and heavy rare earth ions\cite{sarkar2018rare, danmo2023high}. This is different from the current work on (\textit{R}$_1$,...\textit{R}$_n$)VO$_3$ because all \textit{R}VO$_3$ members crystallize into the orthorhombic perovskite structure with the space group \textit{Pbnm} and one does not expect the configurational entropy could play a significant role in forming the solid solution. Second, the size variance might be able to reduce the energy difference between orbital and magnetic order making possible the spin-orbital entanglement in e$_g$ electron systems.  Given the larger energy difference between orbital and spin ordering for manganite perovskites, a larger size variance might be needed to maintain the orbital degeneracy to sufficiently low temperatures for the Kugel-Khomskii spin-orbital interactions. However, the effects of the size variance would be enhanced in manganites  due to the stronger electron-lattice coupling for e$_g$-electrons.

\section{Summary}
In summary, utilizing \textit{R}VO$_3$ as a platform, we investigated the impact of extreme disorder on the orbital degree of freedom within high entropy oxides housing multiple rare earth ions in the same crystallographic site. The findings indicate that the number and type of rare earth ions do not serve as decisive factors in spin and orbital ordering. Instead, two crucial factors are the average ionic radius and size variance, which influence both spin and orbital order. When the size variance is minimal, the average ionic radius dictates the spin and orbital order. Increasing size variance suppresses the long range spin (C-AF) and orbital order (G-OO) but favors the C-OO/G-AF state and the spin-orbital entanglement. The results suggest that the large size variance resulting from mixing multiple rare earth ions can be employed to design a Kugel-Khomskii compound by preserving the  orbital degree of freedom to near T$_N$. \textit{R}VO$_3$ investigated in this work is a model system to study the behavior of t$_{2g}$ electrons. It would be interesting to investigate the spin and orbital order in (\textit{R}$_1$,...,\textit{R}$_n$)MnO$_3$ which provides a platform to investigate the response of e$_g$ orbitals to the lattice disorder in high entropy oxides. Furthermore, the configurational entropy can play an important role in the synthesis of phase pure orthorhombic or hexagonal manganites when mixing light and heavy rare earth ions. The large size variance might lead to spin-orbital entanglement in manganite perovskites or fine tune, for example, the multiferroic properties of hexagonal manganites.

\section{Acknowledgments}
JQY would thank Takeshi Egami, Daniel Khomskii, and Jianshi Zhou for helpful discussions. Work at ORNL  was supported by the U.S. Department of Energy, Office of Science, Basic Energy Sciences, Division of Materials Sciences and Engineering.

\section{references}

%

\end{document}